\documentclass[aps,prl,twocolumn,showpacs]{revtex4}

\usepackage{graphicx}
\usepackage{amsmath,amssymb}
\usepackage{mathtools}
\usepackage{xcolor}
\allowdisplaybreaks

\begin{document}
	
\title{Deciphering mRNA Sequence Determinants of Protein Production Rate}

\author{Juraj Szavits-Nossan}
\email{jszavits@staffmail.ed.ac.uk}
\affiliation{SUPA, School of Physics and Astronomy, University of Edinburgh, Peter Guthrie Tait Road, Edinburgh EH9 3FD, United Kingdom}

\author{Luca Ciandrini}
\affiliation{DIMNP UMR 5235, Universit\'{e} de Montpellier and CNRS, F-34095, Montpellier, France}
\affiliation{Laboratoire Charles Coulomb UMR5221, Universit\'{e} de Montpellier and CNRS, F-34095, Montpellier, France}

\author{M. Carmen Romano}
\affiliation{SUPA, Institute for Complex Systems and Mathematical Biology, Department of Physics,  Aberdeen AB24 3UE, United Kingdom}
\affiliation{Institute of Medical Sciences, University of Aberdeen, Foresterhill, Aberdeen AB24 3FX, United Kingdom}

\begin{abstract}
One of the greatest challenges in biophysical models of translation is to identify coding sequences features that affect the rate of translation and therefore the overall protein production in the cell. We propose an analytic method to solve a translation model based on the inhomogeneous totally asymmetric simple exclusion process, which allows us to unveil simple design principles of nucleotide sequences determining protein production rates. Our solution shows an excellent agreement when compared to numerical genome-wide simulations of \textit{S. cerevisiae} transcript sequences and predicts that the first 10 codons, which is the ribosome footprint length on the mRNA, together with the value of the initiation rate, are the main determinants of protein production rate under physiological conditions. Finally, we interpret the obtained analytic results based on the evolutionary role of codons' choice for regulating translation rates and ribosome densities.
\end{abstract}

\pacs{87.16.aj, 87.10.Mn, 05.60.-k}

% 02. Mathematical methods in physics
%% 02.50.Ga 	Markov processes

% 05. Statistical physics, thermodynamics, and nonlinear dynamical systems
%% 05.60.-k 	Transport processes

% 87. Biological and medical physics
%% 87.10.-e 	General theory and mathematical aspects
%%% 87.10.Hk 	Lattice models
%%% 87.10.Kn 	Finite element calculations
%%% 87.10.Mn 	Stochastic modeling

%% 87.16.-b 	Subcellular structure and processes
%%% 87.16.A- 	Theory, modeling, and simulations
%%% 87.16.ad 	Analytical theories
%%% 87.16.aj 	Lattice models

\maketitle

%--------------------------------------------------------------------------
% INTRODUCTION
%--------------------------------------------------------------------------

Translation is one of the major steps in protein biosynthesis. During this process, the nucleotide sequence of a messenger RNA (mRNA) is translated into a functional protein. Each nucleotide triplet, called codon, codes for a specific amino acid, the proteins' building block. There is experimental evidence that the rate at which a certain mRNA is translated depends on its specific codon sequence~\cite{Gingold11,Kemp13,Chu14,Gorgoni16},  especially in the case of eukaryotes. Identifying sequence features that determine protein production rate, also commonly referred to as translation rate or efficiency, is a fundamental open question in molecular biology~\cite{Gingold11,Chu14}.
	
\begin{figure}[!hbt]
\centering\includegraphics[width=8.6cm]{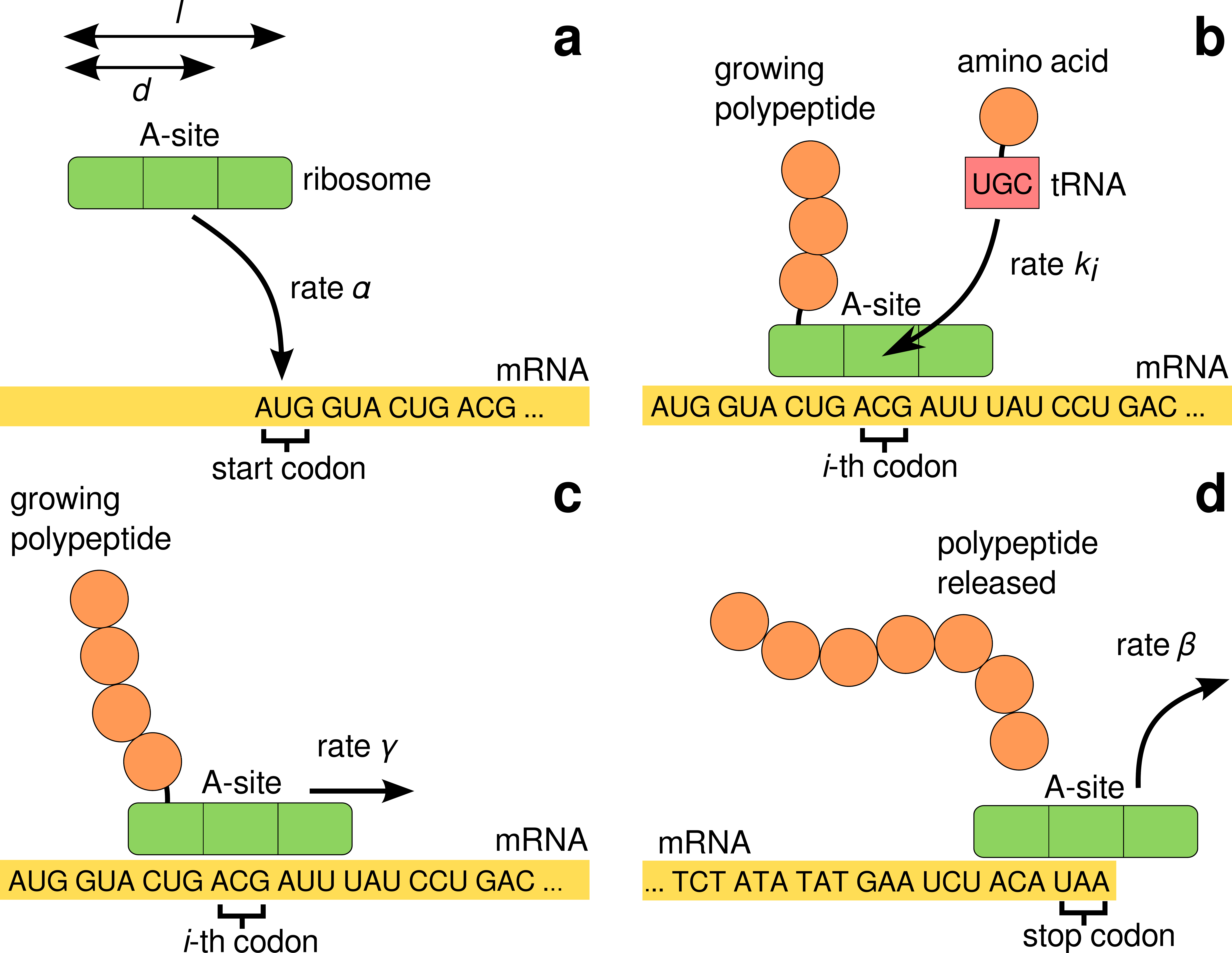}
\caption{Sketch of the mRNA translation process involving initiation (a), elongation consisting of tRNA delivery (b) and translocation (c), followed by termination (d). We emphasise that this is an oversimplified scheme of the process and that actual ribosomes cover $\ell=10$ codons. At each elongation step, the ribosome receives an amino acid from a tRNA that matches the codon occupied by the A-site of the ribosome (the ``reading'' site). After the amino acid is added to the growing polypeptide, the ribosome translocates one codon forward, and the process is repeated.}
\label{fig1}
\end{figure}
	
Translation is performed by molecular motors called ribosomes, which move unidirectionally along the mRNA. The amino acids are delivered to the ribosome by molecules called transfer RNAs (tRNAs), which are specific to the codon and the amino acid they deliver (Fig.~\ref{fig1}). The dwelling time of a ribosome on a specific codon depends primarily on the abundance of the corresponding tRNA~\cite{Varenne84,Sorensen89}. Understanding how codon sequences determine protein production rates can potentially unlock many synthetic applications~\cite{Gorochowski16,Boel16}.
	
The standard biophysical model of translation is known as the totally asymmetric simple exclusion process (TASEP), which captures the concurrent motion of ribosomes on the mRNA~\cite{MacDonald68,Tuller16}. In this model ribosomes progress along the mRNA codon by codon, provided that the codon a ribosome moves onto is not occupied by another ribosome. The protein production rate can then be identified as the ribosomal current of this driven lattice gas. Due to the net current, TASEP is not in equilibrium and its steady state is only known for a few special cases~\cite{DEHP,BlytheEvans07}. Unfortunately, most biologically relevant variants of TASEP can be studied only numerically~\cite{ChouLakatos04,Shaw04,Zia11}, and efficient methods of exploring a large number of parameters are lacking. The steady state of the TASEP with non-uniform hopping rates is a long outstanding problem in nonequilibrium statistical physics~\cite{Zia11,Schmidt15}. An analytic prediction of protein production rates is also needed in order to interpret recently developed ribosome profiling experiments that are capable of monitoring ribosome positions along the mRNA~\cite{Ingolia09}. Such analytical prediction is missing in previous models of ribosome dynamics implicitly or explicitly based on the TASEP~\cite{Gilchrist06, Mitarai08, Reuveni11, Ciandrini13, Shah13}, and it is key to decipher sequence determinants of protein production rates.
	
In this Letter, we develop a versatile analytic method to solve TASEP-based models of translation. Our analytic approach, integrated with simulations and experimental data, allows us to efficiently identify the main features of mRNA codon sequence that determine the rate of protein production. The genome-wide comparison of our analytic predictions with numerical simulations shows an excellent agreement for the model organism \textit{S. cerevisiae} (baker's yeast).  
	
%--------------------------------------------------------------------------
% STOCHASTIC MODEL OF mRNA TRANSLATION
%--------------------------------------------------------------------------

\textit{Stochastic model of mRNA translation.} We focus on the model for translation introduced in~\cite{Ciandrini10,Ciandrini13,Klumpp08}. The mRNA is represented by a one-dimensional lattice consisting of $L$ discrete sites (codons), where site $1$ designates the start codon. Ribosomes are represented by particles that occupy $\ell = 10$ lattice sites, which is the ribosome footprint length measured in ribosome profiling experiments ~\cite{Ingolia09}. We identify the position $1\leq i\leq L$ of a ribosome with the position of its A-site, which is located $d = 5$ lattice sites from the trailing end of the ribosome (Fig.~\ref{fig1}a). A ribosome that waits for a tRNA at position $i$ is labelled by $1_i$ and a ribosome that has already received the correct tRNA and is ready to move is labelled by $2_i$. The set of labels of all translating ribosomes on the lattice is called a configuration $C$ of the system. For example, $C=1_1 2_{13}$ denotes a lattice with 2 ribosomes, one at site $1$ waiting for a tRNA and another one at site $13$ that has already received the correct tRNA.

Ribosomes initiate translation at rate $\alpha$ by binding to the mRNA so that their A-site is at the start codon, provided the sites $1,\dots,\ell-d+1$ are empty (Fig. \ref{fig1}a). A ribosome at site $i$ makes the transition $1_i \rightarrow 2_i$ at rate $k_i$ dependent on tRNA abundances (Fig. \ref{fig1}b) and moves one site forward at rate $\gamma$ (Fig. \ref{fig1}c). Due to steric interactions between particles, the particle at site $i$ can move only if there is no ribosome at site $i+\ell$. Termination occurs when the particle at site $L$ receives the last amino acid, releases the final protein and detaches from the mRNA, which we integrate into a single step occurring at rate $\beta$ (Fig. \ref{fig1}d). 

Our main goal is to compute the rate of protein production as a function of the parameters of the model, which are $\alpha$, $\beta$, $\gamma$ and  $k_i$ for each of the $L$ codons used. We assume that translation takes place under steady-state conditions, so that the ribosomal current is constant along the mRNA and is equal to the rate of protein production, which in turn is equal to rate at which ribosomes load onto the mRNA and initiate translation. To this end we define the codon occupation number $\tau_i$ to be equal to 1 if the $i$-th codon is occupied by an A-site, and $0$ if otherwise. By definition, the exact steady-state ribosomal current $J$ then reads
\begin{equation}
	\label{current}
	J=\alpha\sum_{C}\left[\prod_{i=1}^{\ell}\left(1-\tau_i(C)\right)\right]P(C),
\end{equation}
where $\tau_i(C)$ denotes the $i$-th codon occupation number for configuration $C$, $P(C)$ denotes the steady-state probability that the lattice is in configuration $C$, and the summation goes over all configurations $C$. Other quantities of interest that we compute are the local and total particle densities $\rho_i=\langle\tau_i\rangle$ and $\rho=(1/L)\sum_{i=1}^{L}\rho_i$, respectively.
%
%--------------------------------------------------------------------------
% PERTURBATIVE METHOD FOR THE STEADY STATE
%--------------------------------------------------------------------------

\textit{Series expansion method for computing $P(C)$.} In order to find $P(C)$ one has to solve the steady-state master equation $M\mathbf{P}=0$, where $\mathbf{P}$ is a column vector whose $\mathcal{N}$ elements are the steady-state probabilities $P(C)$ of being in configuration $C$, and $\mathcal{N}$ denotes the total number of configurations. The transition rate matrix $M_{C,C'}$ is given by $W_{C'\rightarrow C}$ for $C\neq C'$ and $-e(C)$ for $C=C'$, where $W_{C'\rightarrow C}$ is the transition rate from $C'$ to $C$ and $e(C)=\sum_{C''\neq C}W_{C\rightarrow C''}$ is the total exit rate from $C$. The exact solution of the master equation can be formally written as
\begin{equation}
	P(C)=\frac{\textrm{det}M^{(p,p)}}{\sum_{q=1}^{\mathcal{N}}\textrm{det}M^{(q,q)}},
	\label{eq:sol_P}
\end{equation}
where $p$ is the position of configuration $C$ in the column vector $\mathbf{P}$ and $\textrm{det}M^{(p,p)}$ is a determinant of the matrix obtained by removing $p$-th row and $p$-th column from $M$ (see Supplemental Material for details). Unfortunately, calculating this determinant is feasible only for unrealistically small system sizes.

To circumvent this problem, we exploit the fact that $\textrm{det}M^{(p,p)}$ is a multivariate polynomial in the variables $\alpha$, $k_1,\dots,k_L$, $\beta$ and $\gamma$~\cite{SzavitsNossan13}. In the biological literature it is often assumed that the initiation rate $\alpha$ is a major limiting step of the translation process, mainly determined by the presence of secondary structures~\cite{Kudla09, Salis09}. For this reason we use $\alpha$ as an expansion parameter and we assume that $\alpha\ll \gamma,\beta,k_1,\dots,k_L$. By collecting terms with the same power of $\alpha$, we can rewrite $P(C)$ as an univariate polynomial $f(C)$ in the variable $\alpha$ with unknown coefficients $f_{n}(C)$
\begin{equation}
	\label{polynomial}
	P(C)=\frac{f(C)}{\sum_{C}f(C)},\quad f(C)=\sum_{n=0}^{K(C)}f_{n}(C)\alpha^n,
\end{equation}
where the coefficients $f_{n}(C)$ depend on the transition rates $k_1,\dots,k_L,\gamma$ and $\beta$ (in order to ease the notation, we leave out the explicit dependence on those parameters). Since we expect that $f(C)$ can be well approximated by the first few terms, the value of $K(C)$ in Eq.~(\ref{polynomial} is irrelevant in our study.

In order to find the unknown coefficients $f_{n}(C)$, we insert Eq.~(\ref{polynomial}) into the master equation $M\mathbf{P}=0$, collect all the terms with the same $n$-th power of $\alpha$ and equate their sum to zero. For a given power $n$, the resulting equation is similar to the original master equation in which $P(C)$ is replaced by $f_{n}(C)$ unless the coefficient multiplying $P(C)$ is $\alpha$, in which case $P(C)$ is replaced by $f_{n-1}(C)$. Starting with $n=0$, we note the equations for the coefficients $f_{0}(C)$ have the same form as the original master equation, but with $\alpha=0$, leading to the trivial solution $f_{0}(C)=0$ if $C\neq\emptyset$. Since the terms $f(C)$ are not normalized, we have the freedom to choose any value for $ f_{0}(\emptyset)$, which we set to $1$.

For $n\geq 1$, the equation for $f_{n}(C)$ reads
\begin{align}
	\label{fn_master}
	& e_0(C)f_{n}(C)+f_{n-1}(C)\sum_{C'}I_{C,C'}=\sum_{C'}I_{C',C}f_{n-1}(C')\nonumber\\
	&\quad+\sum_{C'}(1-I_{C',C})W_{C'\rightarrow C}f_{n}(C')\,,
\end{align}
where $I_{C,C'}=1$ if $W_{C\rightarrow C'}=\alpha$ and is $0$ otherwise, and  $e_0(C)=\sum_{C'}(1-I_{C,C'})W_{C\rightarrow C'}$ is the total exit rate from $C$ excluding the rate $\alpha$. 

A key observation in our analysis is that $f_{n}(C)=0$ whenever the number of particles in $C$ is larger than $n$. This follows from the result $f_{0}(C)= 1 \, (0)$ if $C=\emptyset$ ($C\neq\emptyset$) in conjunction with the hierarchical structure of Eq.~(\ref{fn_master}), which connects configurations differing in the number of particles by no more than one (a proof for $n=1$ is presented in the Supplemental Material). This allows us to write Eq. (\ref{fn_master}) taking into account configurations with only one particle and discarding all the others, which yields
\begin{subequations}
	\label{first-order-eq}
	\begin{align}
		& f_{1}(1_1)=\frac{1}{k_1}f_{0}(\emptyset),\quad f_{1}(1_L)=\frac{\beta}{k_L}f_{1}(2_L)\\
		& f_{1}(1_i)=\frac{\gamma}{k_i} f_{1}(2_{i-1}),\quad 2 \leq i\leq L,\\
		& f_{1}(2_i)=\frac{k_i}{\gamma} f_{1}(1_i),\quad 1\leq i\leq L-1.
	\end{align}
\end{subequations}
The solution to Eqs.~(\ref{first-order-eq}) is given by
\begin{equation}
	\label{f1}
	f_{1}(1_i)=\frac{1}{k_i},\quad f_{1}(2_i)=\begin{cases}\frac{1}{\gamma} & i=1,\dots,L-1\\ \frac{1}{\beta} & i=L\end{cases}
\end{equation}
The equations for $f_{2}(C)$ involving configurations with one and two particles are presented in the Supplemental Material. 

Once we determine the coefficients $f_{n}(C)$ up to a desired order $n$, we can compute the steady-state average of any observable $\mathcal{O}(C)$ by inserting $P(C)$ from Eq.~(\ref{polynomial}) and expanding $\langle\mathcal{O}\rangle$ around $\alpha=0$,
\begin{equation}
	\label{series}
	\langle \mathcal{O}(C)\rangle=\frac{\sum_{n=0}^{K}\sum_{C}\mathcal{O}(C)f_{n}(C)\alpha^n}{\sum_{n=0}^{K}\sum_{C}f_{n}(C) \alpha^n}=\sum_{n=0}^{\infty}c_n\alpha^n.
\end{equation}
For example, the first three coefficients $c_0$, $c_1$ and $c_2$ are given by
\begin{equation}
	c_0=\frac{a_0}{b_0},\enskip c_1=\frac{a_1-c_0 b_1}{b_0},\enskip c_2=\frac{a_2-c_1 b_1-c_0 b_2}{b_0},
\end{equation}
where $a_n$ and $b_n$ are defined as
\begin{equation}
	a_n=\sum_{C}\mathcal{O}(C)f_{n}(C), \quad b_n=\sum_{C}f_{n}(C).
\end{equation}
Notice that the expansion in Eq. (\ref{series}) is slightly different for the current $J$ due to an extra $\alpha$ in Eq. (\ref{current}) and is given by $J=\sum_{n=0}^{\infty}c_n \alpha^{n+1}$. Using the expressions for $f_{n}(C)$ and $f_{n}(C)$ computed earlier yields 
\begin{align}
	& J=\alpha\left[1-\sum_{i=1}^{\ell}\left(\frac{1}{k_i}+\frac{1}{\gamma_i}\right)\alpha+O(\alpha^2)\right],\label{j1}\\
	& \rho=\frac{1}{L}\sum_{i=1}^{L}\left(\frac{1}{k_i}+\frac{1}{\gamma_i}\right)\alpha+O(\alpha^2),\label{rho1}\\
	& \rho_i= \left(\frac{1}{k_i}+\frac{1}{\gamma_i}\right)\alpha+O(\alpha^2),\label{tau1}
\end{align}
where $\gamma_i=\gamma+(\beta-\gamma)\delta_{i,L}$. These equations constitute our main result. Equation~(\ref{j1}) shows that if the initiation rate $\alpha$ is small compared to $k_1,\dots,k_L$ and $\gamma$, than the protein production rate $J$ depends predominately on the initiation rate, along with the translocation and elongation rates of the first 10 codons, corresponding to the ribosome footprint $\ell$. The importance of the first 10 codons is a direct consequence of the excluded volume interactions: any ribosome already present in that region will prevent a new ribosome from binding the mRNA. It is also important to emphasize that (\ref{j1})--(\ref{tau1}) are exact series expansions around $\alpha=0$; the approximation is made only when the series is truncated.

\textit{Independent Particle Approximation (IPA).} Interestingly, the excluded volume interactions between particles have no effect on the first two terms in the series expansion of $J$. This motivates us to ask how the expansion in Eq.~(\ref{j1}) would look like if we assumed that all particles are independent, i.e. not experiencing any exclusion interaction. In our model, the IPA amounts to replacing $P(C)$ with
\begin{equation}
	\label{IPA}
	P^{\textrm{IPA}}(C)=\frac{1}{Z_L}\prod_{j=1}^{N(C)}w_{1}(\theta_{X(j)}),
\end{equation}
where $N(C)$ is the number of particles in a configuration $C$, $\theta$ is one of the two particle states $1$ and $2$, $X(i)$ is the position of the $i$-th particle on the lattice and $Z_L$ is the normalization constant. The weights $w_{1}(1_i)$ and $w_{1}(2_i)$ for $i=1,\dots,L$ are obtained by solving the master equation for a single particle and are given by $\alpha/k_i$ and $\alpha/\gamma_i$, respectively. The corresponding expressions for the current $J$ and local density $\rho_i$ read
\begin{equation}
	J^{\textrm{IPA}}=\frac{\alpha}{\prod_{i=1}^{\ell}(1+p_i)},\quad \rho^{\textrm{IPA}}_{i}=\frac{p_i}{1+p_i},
\end{equation}
where $p_i=\alpha(1/k_i+1/\gamma_i)$. We will use these results later in order to determine the importance of ribosome collisions in real genetic sequences. We also note that the IPA in (\ref{IPA}) provides a good approximation to $f_{n}(C)$ for $n=N(C)$ when the particles in $C$ are far apart from each other.

%--------------------------------------------------------------------------
% mRNA TRANSLATION IN YEAST
%--------------------------------------------------------------------------

\textit{Application to mRNA translation in yeast.} We now apply our results to the transcriptome of \textit{S. cerevisiae} using realistic model parameters. The values of $\alpha$ in the range $0.005-4.2$ s$^{-1}$ with the median value of $0.09$ $s^{-1}$ have been previously estimated in Ref.~\cite{Ciandrini13} using genome-wide experimental values of the ribosomal density. We assume that the rates $k_i$ are mainly proportional to the gene copy number of tRNAs delivering the corresponding amino acid~\cite{Percudani97}; the rates are normalized so that the average codon translation rate is equal to the experimental value of $10$ codons/s~\cite{Arava03}; the estimates of all elongation rates along with the distribution of $\alpha$ is presented in the Supplemental Material. The translocation rate $\gamma$ is fixed to $\gamma=35$ codons/s~\cite{Savelsbergh03}, and termination is assumed to be fast and comparable to translocation~\cite{Arava03}, $\beta\approx\gamma$, so that $\gamma_i=\gamma \;\forall i$ in Eqs. (\ref{j1})-(\ref{tau1}). 

In total, we analyzed 5836 gene sequences; for each gene we calculated $J$, $\rho$ and $\rho_i$ for $1\leq i\leq L$ up to and including the second order of the perturbative expansion at the corresponding physiological value of $\alpha$. The results were then compared to the exact values obtained numerically with stochastic simulations using the Gillespie algorithm~\cite{Gillespie77} by calculating the percent error $\epsilon$.

\begin{figure}[hbt]
	\centering\includegraphics[width=8.6cm]{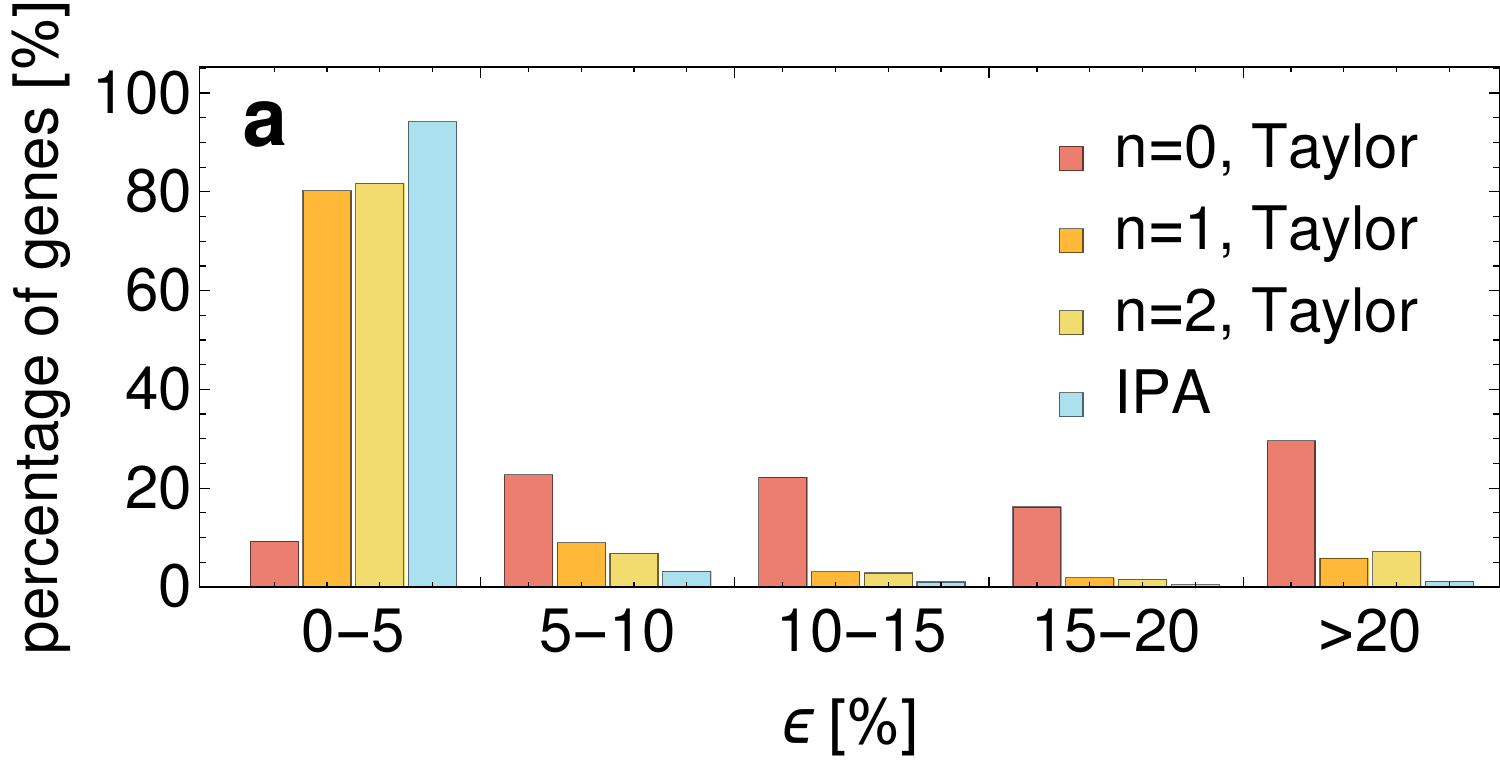}\\
	\centering\includegraphics[width=8.6cm]{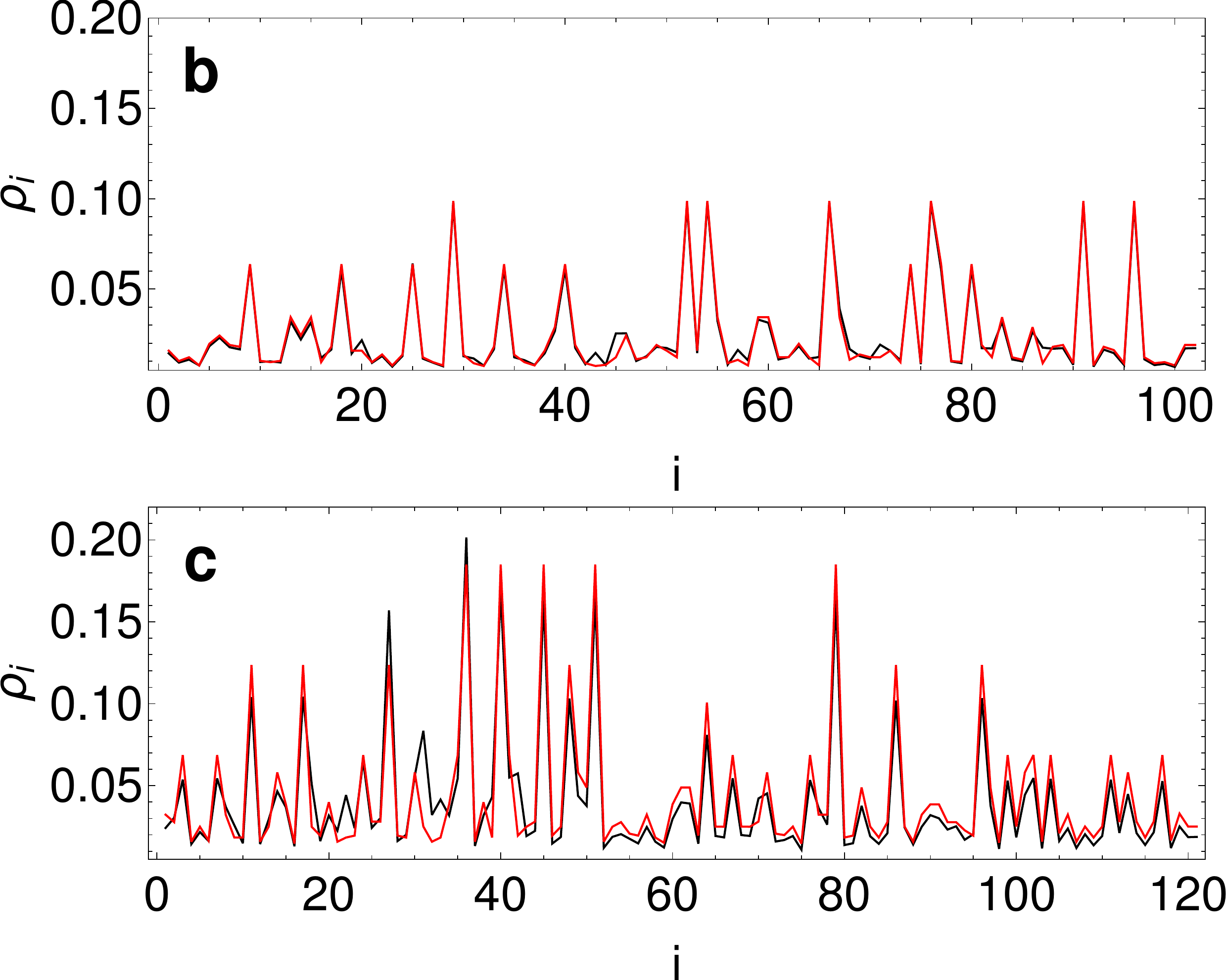}
	\caption{(a): Histogram of the percent error $\epsilon$ measuring the discrepancy between the protein production rate obtained by stochastic simulations and the perturbative expansion including the independent-particle approximation (IPA). (b) and (c): Ribosomal density profile obtained by stochastic simulations (black) compared to Eq. (\ref{tau1}) (red) for two values of the initiation rate $\alpha$, one close to the median value $\approx0.09$ s$^{-1}$ (gene YAL045C, b) and the other that is close to the 90th percentile value $\approx 0.23$ s$^{-1}$ (gene YGL034C, c).}
	\label{fig2}
\end{figure}

For the protein production rate $J$, with the zeroth order of the perturbative expansion (predicting that $J=\alpha$) we obtain an error of $\epsilon < 5\%$ for only $11\%$ of the genes. Remarkably, that percentage jumps to $80.7\%$ when the first-order coefficients in Eq.~(\ref{f1}) are taken into account (Fig.~\ref{fig2}). Including the second-order coefficients (computed numerically from Eq.~(\ref{fn_master})) does not significantly improve results, due to a large value of $\alpha\gtrapprox 0.15$ s$^{-1}$ in about $20\%$ of the genes. Since the coefficients in Eq. (\ref{series}) typically alternate in sign, truncating the series will ultimately lead to a wrong result when the value of $\alpha$ is large enough. On the other hand, the IPA does not suffer from this problem and leads to an error $\epsilon<5\%$ in $94\%$ of genes, whereby only $1\%$ of genes have $\epsilon>20\%$. The success of the IPA also suggests that ribosome collisions and traffic jams have a minor effect on the rate of translation, which is in accordance with recent experimental evidence~\cite{Laursen10,Kudla09}. This is also apparent from the density profile $\rho_i$, which is very well approximated by the linear approximation in Eq.~(\ref{tau1}), even for larger values of $\alpha$ (Fig. \ref{fig2} b and c).

%--------------------------------------------------------------------------
% IDENTIFYING DETERMINANTS OF mRNA TRANSLATION I
%--------------------------------------------------------------------------

\textit{Identifying determinants of mRNA translation.} Our analytic prediction allows us to decompose the contributions from initiation and elongation to the rate of translation $J$, thereby addressing a long-standing question about main determinants of protein production rate. Remarkably, the expressions for the current obtained with both the first-order and the independent particle approximation involve only the first $\ell=10$ codons. This result therefore strongly indicates that, together with the initiation rate, the first 10 codons of the mRNA are the key determinants of the protein production rate by preventing a new ribosome from binding which effectively decreases the initiation rate.

\begin{figure}[!htb]
	\centering\includegraphics[width=8.6cm]{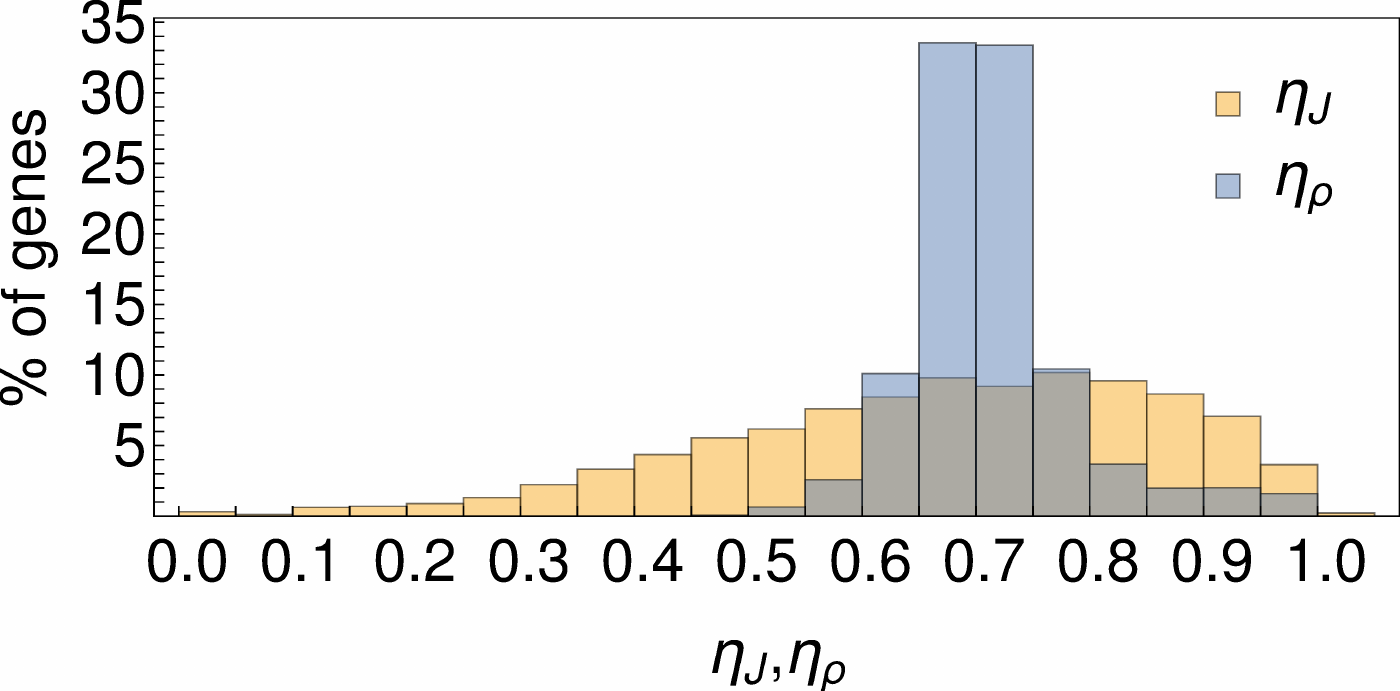}
	\caption{Histogram of $\eta_J$ (yellow) and $\eta_\rho$ (blue) for the \textit{S. cerevisiae} genome.}
	\label{fig3}
\end{figure}

If we assume that the cell maximizes the rate of protein production, given the above result, we would expect to find a signature in the genome for selecting efficient \textit{fast} codons at the beginning of each gene. We test this hypothesis by computing  the ribosomal current for the fastest ($J^{F}$) and the slowest ($J^{S}$) set of first $\ell$ synonymous codons for each gene, using the IPA. A score $\eta_J=(J-J^{S})/(J^{F}-J^{S})$ is then assigned to each gene, which is $1$ ($0$) when the sequence corresponds to the fastest (slowest) codon sequence. On the other hand, one might assume that the cell not only tries to maximize protein production rates, but at the same time it tries to minimize the ribosome density $\rho$ on mRNAs. This assumption is motivated by the fact that ribosomes are limiting~\cite{Shah13, Greulich12} and highly costly in terms of cellular energy resources~\cite{Keiler15}, and therefore ribosome queues are to be avoided. Hence, we also compute $\eta_\rho=(\rho-\rho^S)/(\rho^F-\rho^S)$ for each gene, where $\rho^F$ and $\rho^S$ denote the ribosome density for the fastest and slowest set of synonymous codons, respectively.

Figure~\ref{fig3} shows the histogram of $\eta_{J}$ (yellow) and $\eta_\rho$ (blue) computed for 5836 genes of \textit{S.cerevisiae}. Both histograms show an average of $0.7$ suggesting the selection of fast codons near the start codon to maximize $J$, as well as an overall selection of fast codons along the mRNA to minimize $\rho$. However, the width of the distribution of $\eta_\rho$ is substantially smaller than the one of $\eta_J$. This might indicate that the optimization of protein production rate is strongly dependent on the particular gene, since different proteins are needed at different concentrations. In contrast, the minimization of the number of ribosomes on mRNAs could be a more general constraint. 

%--------------------------------------------------------------------------
% CONCLUSIONS
%--------------------------------------------------------------------------

\textit{Conclusions.} We have presented an analytic method that allows us to quantify sequence determinants of protein production rates, using a model for translation that is based on an inhomogeneous exclusion process. Our results demonstrate that the rate of protein production is largely determined by the initiation rate and the elongation rates of the first 10 codons (the ribosome footprint length on the mRNA), which control how fast ribosomes load onto the mRNA, whereby ribosome collisions and queues have a minor effect under physiological conditions.

%--------------------------------------------------------------------------
% ACKNOWLEDGMENTS
%--------------------------------------------------------------------------

\begin{acknowledgments}
\textit{Acknowledgments.} MCR and LC contributed equally to this work. JSN was supported by the Leverhulme Trust Early Career Fellowship. MCR was supported by the Biotechnology and Biological Sciences Research Council (BBSRC) BB/N017161/1 and the Scottish Universities Life Sciences Alliance. LC thanks the CNRS for having granted him a “demi-d\'{e}́l\'{e}gation” (2017--18).
\end{acknowledgments}

%--------------------------------------------------------------------------
% REFERENCES
%--------------------------------------------------------------------------

%--------------------------------------------------------------------------
% SUPPLEMENTAL MATERIAL
%--------------------------------------------------------------------------

\onecolumngrid
\newpage

%\maketitle
\begin{center}
	{\Large \bf  Supplemental Material to:\\ Deciphering mRNA Sequence Determinants of \\ Protein Production Rate}
\end{center}
\medskip

\allowdisplaybreaks[1]

\setcounter{equation}{0}

\setcounter{figure}{0}
\renewcommand{\thefigure}{S\arabic{figure}}
\renewcommand{\theequation}{S\arabic{equation}}

%------------------------------------------------
% EXACT SOLUTION
%------------------------------------------------

\section{Exact solution of the steady-state master equation}

The exact solution, Eq. (3) in the main text, pertains to any ergodic Markov jump process with a finite number of states. In order to derive the exact solution, let us write again the steady-state master equation in a matrix form
\begin{equation}
	\label{master}
	M\mathbf{P}=0,
\end{equation}
where $\mathbf{P}$ is a column vector whose elements are steady-state probabilities $P_i$, the $\mathcal{N}$ states are indexed by $i=1\dots,\mathcal{N}$ and the $\mathcal{N}\times \mathcal{N}$ matrix $M$ is given by
\begin{equation}
	\label{M}
	M_{ij}=\begin{cases}
		W_{jk} & i\neq j\\
		-\sum_{k\neq i}W_{ik} & i=j,
	\end{cases}
\end{equation}
where $W_{ij}$ is a transition rate from a state $i$ to $j$. We note from Eq. (\ref{M}) that the sum of all elements in each column of $M$ is zero
\begin{equation}
	\sum_{i=1}^{\mathcal{N}}M_{ij}=\sum_{i=1 \atop i\neq j}^{\mathcal{N}}W_{ji}-\sum_{k=1 \atop k\neq j}^{\mathcal{N}}W_{jk}=0.
\end{equation}
Consequently, the sum all row vectors of $M$ is zero, which means that the vectors are linearly dependent and thus the determinant of $M$ is equal to zero,  
\begin{equation}
	\label{detm}
	\textrm{det}M=0.
\end{equation} 
Combining Eq. (\ref{detm}) with the Laplace expansion of a determinant yields
\begin{equation}
	\label{laplace}
	0=\textrm{det}M=\sum_{i=1}^{\mathcal{N}}M_{ij}C_{ij}=\sum_{j=1}^{\mathcal{N}}M_{ij}C_{ij}.
\end{equation}
Here $C_{ij}$ is a cofactor of $M$ defined as $C_{ij}=(-1)^{i+j}\textrm{det}M^{(i,j)}$, where the matrix $M^{(i,j)}$ is obtained from $M$ by removing $i$-th row and $j$-th column . Inserting Eq. (\ref{M}) into (\ref{laplace}) gives
\begin{align}
	\textrm{det}M&=\sum_{i=1 \atop i\neq j}^{\mathcal{N}}M_{ij}C_{ij}+M_{jj}C_{jj}=\sum_{i=1 \atop i\neq j}^{\mathcal{N}}M_{ij}C_{ij}-\sum_{k=1 \atop k\neq j}^{\mathcal{N}}M_{kj}C_{jj}\nonumber\\
	&=\sum_{i=1 \atop i\neq j}^{\mathcal{N}}M_{ij}(C_{ij}-C_{jj})=0,
\end{align}
from which we conclude that 
\begin{equation}
	\label{Cij}
	C_{ij}=C_{jj},
\end{equation}
for any $i$ and $j$, i.e. the cofactor $C_{ij}$ does not depend on the state $i$. Inserting Eq. (\ref{Cij}) back into Eq. (\ref{detm}) gives
\begin{equation}
	\sum_{j=1}^{\mathcal{N}}M_{ij}C_{jj}=0,
\end{equation}
which is precisely the starting steady-state master equation, Eq. (\ref{master}). We thus conclude that
\begin{equation}
	P_{i}=\frac{C_{ii}}{\sum_{j}C_{jj}}=\frac{\textrm{det}M^{(i,i)}}{\sum_{j=1}^{\mathcal{N}}\textrm{det}M^{(j,j)}}.
\end{equation}

\section{Argument for $f^{(n)}(C)=0$ whenever the number of particles in \newline a configuration $C$ is larger than $n$}

As noted in the main text, a key observation in our analysis is that $f^{(n)}(C)=0$ whenever the number of particles in $C$ is larger than $n$. We present here the argument for the case of $n=1$; a similar argument applies to higher orders. For $n=1$, using Eq.~(5) in the main text we obtain
\begin{equation}
	\label{f1_master_b}
	f^{(1)}(C)=\frac{1}{e(C)}\bigg[\sum_{C'}I_{C'\to C}\delta_{C',\emptyset}+\sum_{C'}(1-I_{C'\rightarrow C})W_{C'\rightarrow C}f^{(1)}(C') - \delta_{C,\emptyset}\bigg].
\end{equation}
Consider $C$ to be one full lattice configuration with all $N=\lfloor (L-1+\ell)/\ell\rfloor$ particles in state 1. The first term on the right hand side of Eq.~(\ref{f1_master_b}) is zero because there is no direct transition from the empty configuration $\emptyset$ to $C$. The second term on the right hand side is also zero, because $C$ can only be accessed through the transition rate $\alpha$. Finally, the third term on the right hand side is clearly zero. Hence, $f^{(1)}(C)=0$. Now, it is easily seen that all full lattice configurations $C'$ that can be accessed only from $C$ will fulfil $f^{(1)}(C')=0$. Iterating further this procedure to the configurations that can be accessed from those, it becomes clear that all full lattice configurations $C$ have $f^{(1)}(C)=0$. This argument can be further extended to all configurations with $N-1, N-2, N-3, ..., 2$ particles, resulting in $f^{(1)}(C)=0$ for all of them. 

The situation changes when we consider configurations with only 1 particle on the lattice, since then there is a direct transition from the empty configuration $\emptyset$ to one of those configurations, leading to $f^{(1)}(C) \neq 0$. The rest of the coefficients $f^{(1)}(C)$ for 1-particle configurations will depend on each other, and hence, they do not vanish. Therefore, we can conclude that only 1-particle configurations have non-vanishing $f^{(1)}(C)$. For $n=1$, this key observation allows us to write Eq. (5) taking into account configurations with only one particle and discarding all the others, which yields Eqs. (6a)-(6c) in the main text.

\section{Equations for the second-order coefficients $f^{(2)}(C)$}

As stated in the main text, all second-order coefficients $f^{(2)}(C)$ whereby a configuration C has more than two particles are equal to zero. The equations for the non-zero coefficients $f^{(2)(C)}$ are presented below, where we use the the notation $\gamma_j=\gamma$ for $1\leq j\leq L-1$ and $\gamma_{L}=\beta$.

First, we look at configurations with one particle at site $i=1$ and the other at site $j=i+l,\dots,L$:
\begin{subequations}
	\label{1j}
	\begin{align}
		& f^{(2)}(1_1 1_{j})=\frac{1}{k_1+k_j}\left[f^{(1)}(1_j)+(1-\delta_{1+l,j})\gamma_{j-1}f^{(2)}(1_1 2_{j-1})\right]\\
		& f^{(2)}(1_1 2_{j})=\frac{1}{k_1+\gamma_j}\left[f^{(1)}(2_j)+k_j f^{(2)}(1_1 1_j)\right]\\
		& f^{(2)}(2_1 1_j)=\frac{1}{(1-\delta_{1+l,j})\gamma_1+k_j}\left[k_1 f^{(2)}(1_1 1_j)+(1-\delta_{1+l,j})\gamma_{j-1}f^{(2)}(2_1 2_{j-1})\right]\\
		& f^{(2)}(2_1 2_j)=\frac{1}{(1-\delta_{1+l,j})\gamma_1+\gamma_j}\left[k_1 f^{(2)}(1_1 2_j)+k_j f^{(2)}(2_1 1_j)\right].
	\end{align}
\end{subequations}
where we used the Kronecker delta to account for the excluded volume interaction. Next, we look at configurations with particles at sites $i=2,\dots,L-l$ and $j=i+l,\dots,L$.
\begin{subequations}
	\label{ij}
	\begin{align}
		& f^{(2)}(1_i 1_{j})=\frac{1}{k_i+k_j}\left[\gamma_{i-1}f^{(2)}(2_{i-1} 1_j)+(1-\delta_{i+l,j})\gamma_{j-1}f^{(2)}(1_i 2_{j-1})\right]\\
		& f^{(2)}(1_i 2_{j})=\frac{1}{k_i+\gamma_j}\left[\gamma_{i-1}f^{(2)}(2_{i-1}2_j)+k_j f^{(2)}(1_i 1_j)\right]\\
		& f^{(2)}(2_i 1_j)=\frac{1}{(1-\delta_{i+l,j})\gamma_i+k_j}\left[k_i f^{(2)}(1_i 1_j)+(1-\delta_{i+l,j})\gamma_{j-1}f^{(2)}(2_i 2_{j-1})\right]\\
		& f^{(2)}(2_i 2_j)=\frac{1}{(1-\delta_{i+l,j})\gamma_i+\gamma_j}\left[k_i f^{(2)}(1_i 2_j)+k_j f^{(2)}(2_i 1_j)\right].
	\end{align}
\end{subequations}
Finally, the equations for configurations with only one particle are given by
\begin{subequations}
	\label{s1}
	\begin{align}
		& f^{(2)}(1_1)=\frac{1}{k_1}\left[\gamma_L f^{(2)}(1_1 2_L)+f^{(1)}(\emptyset)\right],\\
		& f^{(2)}(2_1)=\frac{k_1}{\gamma_1}f^{(2)}(1_1)+\frac{\gamma_L}{\gamma_1}f^{(2)}(2_1 2_L),\\
		& f^{(2)}(1_i)=\frac{1}{k_i}\left[\gamma_{i-1}f^{(2)}(2_{i-1})+\theta(L-l+1-i)\gamma_L f^{(2)}(1_i 2_L)-f^{(1)}(1_i)\right],\quad i=2,\dots,L\\
		& f^{(2)}(2_i)=\frac{1}{\gamma_i}\left[k_i f^{(2)}(1_{i})+\theta(L-l+1-i)\gamma_L f^{(2)}(2_i 2_L)-f^{(1)}(2_i)\right],\quad i=2,\dots,L
	\end{align}
\end{subequations}
where $\theta(n)=0$ for $n<0$ and $1$ for $n\geq 0$. 

The equations (\ref{1j})-(\ref{s1}) can be easily solved numerically by iteration. We first solve Eqs. (\ref{1j}) for $i=1,j=l+1$. We then iterate the recursion relation in Eqs. (\ref{1j}) for fixed $i=1$ and $j=2+l,\dots,L$. We then solve equations (\ref{ij}) for fixed $i=2$ and $j=l+3,\dots,L$ and repeat this procedure until $i=L-l$ and $j=L$. Finally, we calculate the coefficients $f^{(2)}(1_i)$ and $f^{(2)}(2_i)$ for the single-particle configurations using $f^{(2)}(1_i 2_L)$ and $f^{(2)}(2_i 2_L)$ that we solved in the previous steps. We note that the only unknown that we cannot determine is $f^{(1)}(\emptyset)$, but it turns out that all terms containing $f^{(1)}(\emptyset)$ will cancel out later when we calculate the coefficients in the series expansions (11)-(13) in the main text.

\section{Estimates of the codon elongation rates}

The total translation rate $\omega_i$ of the codon $i$ is:
\begin{equation}
	\frac{1}{k_i}+\frac{1}{\gamma}=\frac{1}{\omega_i},
\end{equation}
where the translocation rate $\gamma$ is set to $35s^{-1}$ as mentioned in the main text. The codon elongation rate $k_i$, which represents the average arrival and recognition time of the cognate tRNA, is determined by following the procedure introduced in~\cite{Ciandrini13}, which we report here. 

For each of the 41 tRNAs types $j$ we consider their gene copy number (GCN), which allows us to provide a first estimate of the rate $k_j$:
\begin{equation}\label{k}
	k_j=r\frac{GCN_j}{\sum_{j=1}^{41} GCN_j},
\end{equation}
where $GCN_j$ is the gene copy number of the tRNA of type $j$ with $j=1,\dots,41$, and $r$ is a proportionality constant. These rates were then adjusted to take into account experimental evidence suggesting that the translation rates of codons using the G-U wobble are reduced by 39\% compared to their G-C counterparts; analogously, codons using the wobble I-C and codons using the wobble I-A are reduced by 36\% relative to their I-U counterparts~\cite{Gilchrist06}. To calculate the proportionality constant $r$, we used the experimental value of $10$ codons/s for the average codon translation rate $\langle \omega_i \rangle$ defined as
\begin{equation}
	\langle \omega_i \rangle=\sum_{i=1}^{61}\left( \frac{k_i \gamma}{k_i+\gamma}\right)\frac{n_i}{n}
\end{equation}
where $n_i/n$ is the relative abundance of all different codon types in the cell, $n_i$ is total number of codons in the cell of exactly type $i$, and $n=\sum_{i=1}^{61}n_i$ (the three STOP codons are excluded)~\cite{Ciandrini13}. Table~\ref{t1} summarises the resulting elongation rates $k_i$ for 61 codons. 

\begin{table*}[h!]
	\begin{center}
		\begin{tabular}{|c|c|c|c|c|}
			\hline
			tRNA & anti-codon & codon  & GCN & $k_i$ [1/s] \\
			\hline
			\hline
			Ala1&IGC&GCU&11.00&18.34\\
			Ala1&IGC&GCC&11.00&11.74\\
			Ala2&UGC&GCA&5.00&8.34\\
			Ala2&UGC&GCG&5.00&8.34\\
			Arg1&CCU&AGG&1.00&1.67\\
			Arg2&ICG&CGU&6.00&10.01\\
			Arg2&ICG&CGC&6.00&6.40\\
			Arg2&ICG&CGA&6.00&6.40\\
			Arg3&UCU&AGA&11.00&18.34\\
			Arg4&CCG&CGG&1.00&1.67\\
			Asn&GUU&AAU&10.00&10.17\\
			Asn&GUU&AAC&10.00&16.68\\
			Asp&GUC&GAU&15.00&15.26\\
			Asp&GUC&GAC&15.00&25.01\\
			Cys&GCA&UGU&4.00&4.07\\
			Cys&GCA&UGC&4.00&6.67\\
			Gln1&UUG&CAA&9.00&15.01\\
			Gln2&CUG&CAG&1.00&1.67\\
			Glu3&UUC&GAA&14.00&23.35\\
			Glu4&CUC&GAG&2.00&3.34\\
			Gly1&GCC&GGU&16.00&16.28\\
			Gly1&GCC&GGC&16.00&26.68\\
			Gly2&UCC&GGA&3.00&5.00\\
			Gly3&CCC&GGG&2.00&3.34\\
			His&GUG&CAU&7.00&7.12\\
			His&GUG&CAC&7.00&11.67\\
			Ile1&UAU&AUA&2.00&3.34\\
			Ile2&IAU&AUU&13.00&21.68\\
			Ile2&IAU&AUC&13.00&13.87\\
			Leu1&UAG&CUA&3.00&5.00\\
			Leu1&UAG&CUG&3.00&5.00\\
			\hline
		\end{tabular}
		\quad
		\begin{tabular}{|c|c|c|c|c|}
			\hline
			tRNA & anti-codon & codon  & GCN & $k_i$ [1/s]\\
			\hline
			\hline
			Leu3&CAA&UUG&10.00&16.68\\
			Leu4&UAA&UUA&7.00&11.67\\
			Leu5&GAG&CUU&1.00&1.02\\
			Leu5&GAG&CUC&1.00&1.67\\
			Lys1&CUU&AAG&14.00&23.35\\
			Lys2&UUU&AAA&7.00&11.67\\
			Met&CAU&AUG&5.00&8.34\\
			Phe&GAA&UUU&10.00&10.17\\
			Phe&GAA&UUC&10.00&16.68\\
			Pro1&UGG&CCA&10.00&16.68\\
			Pro1&UGG&CCG&10.00&16.68\\
			Pro2&IGG&CCU&2.00&3.34\\
			Pro2&IGG&CCC&2.00&2.13\\
			Ser2&IGA&UCU&11.00&18.34\\
			Ser2&IGA&UCC&11.00&11.74\\
			Ser3&GCU&AGU&4.00&4.07\\
			Ser3&GCU&AGC&4.00&6.67\\
			Ser4&UGA&UCA&3.00&5.00\\
			Ser5&CGA&UCG&1.00&1.67\\
			Thr1&IGU&ACU&11.00&18.34\\
			Thr1&IGU&ACC&11.00&11.74\\
			Thr2&CGU&ACG&1.00&1.67\\
			Thr3&UGU&ACA&4.00&6.67\\
			Trp&CCA&UGG&6.00&10.01\\
			Tyr&GIA&UAU&8.00&8.14\\
			Tyr&GIA&UAC&8.00&13.34\\
			Val1&IAC&GUU&14.00&23.35\\
			Val1&IAC&GUC&14.00&14.94\\
			Val2&UAC&GUA&2.00&3.34\\
			Val2b&CAC&GUG&2.00&3.34\\
			\hline
		\end{tabular}
	\end{center}
	\caption{Elongation rates $k_i$ considering supply (gene copy number of tRNAs) and wobble base-pairing.}
	\label{t1}
\end{table*}%

The estimates provided are valid in physiological conditions, when amino acids are not limiting. In case of amino acid starvation the rates of the codons affected should be modified (the elongation rates depend on the abundance of tRNAs charged with the correct amino acid).

\section{Distribution of the translation initiation rates}

\begin{figure}[!hbt]
	\centering\includegraphics[width=12cm]{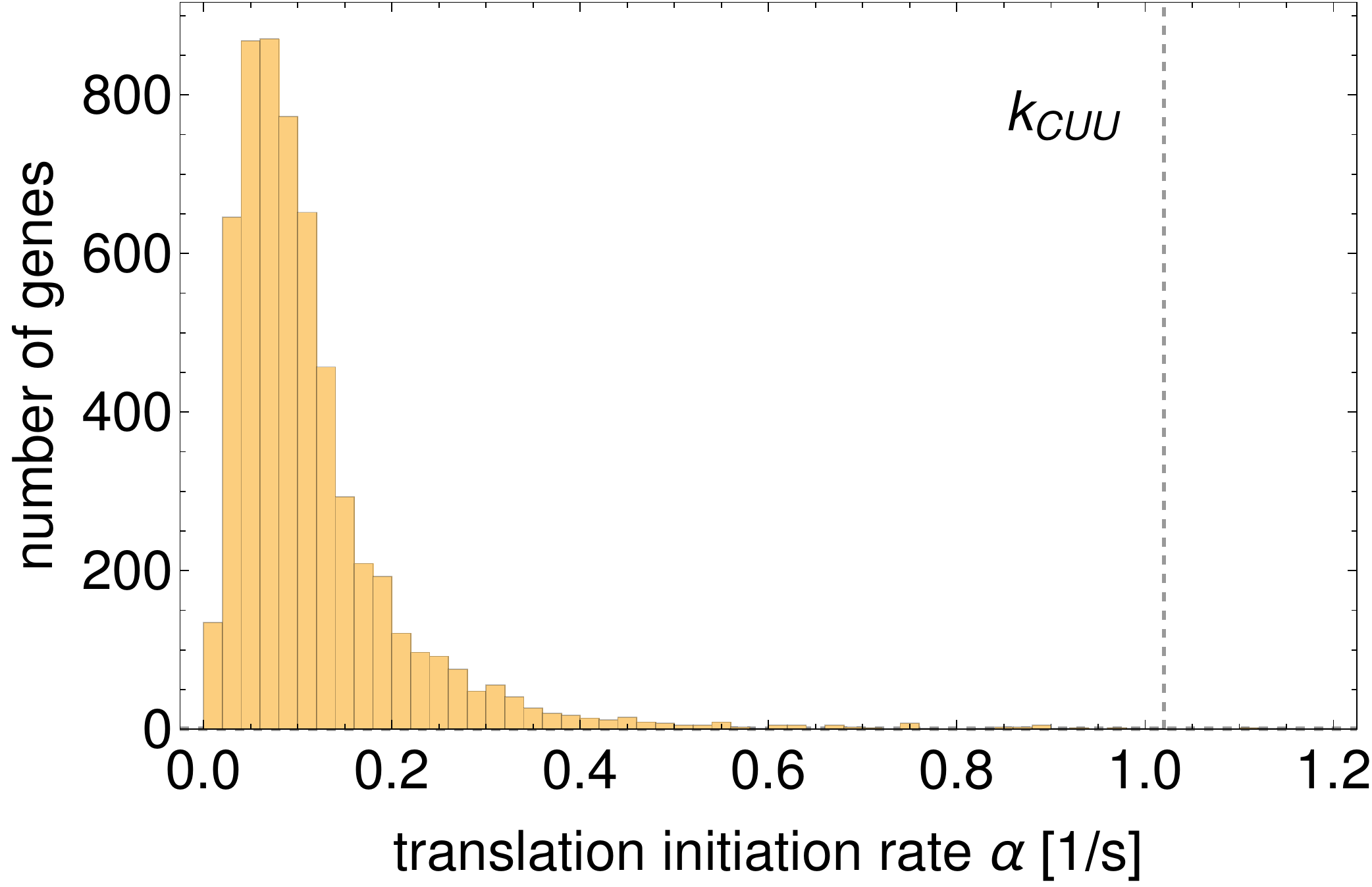}
	\caption{Histogram of the estimated initiation rates for 5836 genes from the \emph{S. cerevisiae} genome. The 95th percentile value of $\alpha$ is $\approx 0.3$, which is much smaller that the smallest elongation rate, which is $1.02$ codons/s for the CUU codon (vertical dashed line). Only 15 genes have $\alpha$ larger than $\textrm{min}_i\{k_i\}=1.02$ codons/s.}
	\label{figS1}
\end{figure}

The translation initiation rates $\alpha$ were estimated for each gene in Ref.~\cite{Ciandrini13} by matching the average ribosome density predicted by the model with the experimental value of the ribosome density obtained in Ref.~\cite{Arava03}. The resulting values for 5836 genes that we analyzed in the main text are in the range $0.005-4.2$ s$^{-1}$ with the median value of $0.09$ $s^{-1}$. The histogram of the rates is presented in Figure \ref{figS1}, showing that the initiation rate $\alpha$ for the majority of genes is indeed much smaller than the smallest elongation rate, which has the value of $1.02$ codons/s for the CUU codon (see Table \ref{t1}). The list of all translation initiation rates can be found in the Supplemental Material of Ref.~\cite{Ciandrini13}.

\end{document}